\documentclass[12pt, draft]{revtex4-2}
\usepackage{amssymb, amsmath}

\begin{document}
\title{ Aspects of thermal one-point functions and response functions in AdS Black holes  }

\author{David Berenstein, Robinson Mancilla}
\address{Department of Physics, University of California, Santa Barbara, CA 93101}

\begin{abstract}
    We revisit the problem of analytically computing the one point functions for scalar fields in planar AdS black holes of arbitrary dimension, which are sourced by the Weyl squared tensor. We analyze the problem in terms of power series expansions around the boundary using the method of Frobenius. We clarify the pole structure of the final answer in terms of operator mixing, as argued previously by Grinberg and  Maldacena. We generalize the techniques to also obtain analytic results for slowly modulated spatially varying sources to first non-trivial order in the wave vector for arbitrary dimension.     We also study the first order corrections to the one point function of the global AdS black hole at large mass, where we perturb in terms that correspond to the curvature of the horizon. 
\end{abstract}

\maketitle

\section{Introduction}

The AdS/CFT correspondence \cite{Maldacena:1997re} provides information about field theories at strong coupling by performing gravity calculations. Correlators of the field theory can be expressed as boundary correlators of the AdS geometry \cite{Gubser:1998bc,Witten:1998qj}.
In particular an AdS black hole corresponds to a thermal state in the field theory. Usually, one uses the eigenstate thermalization hypothesis (ETH) \cite{Deutsch, Srednicki:1994mfb} to argue that it does not matter if the state in question is pure (with the right energy, charge, etc) or mixed: their average properties will be the same.
It is important that if the state in question is pure, some correlation
functions that can be studied  can also be interpreted as OPE coefficients or four point functions of primary operators in conformal field theory, whereas in a mixed state, they are averages of OPE coefficients, or four point functions. The ETH tells us that both results will be almost identical to each other with exponentially suppressed differences.

Non-trivial one point functions in a thermal background can be interpreted as OPE coefficients if we work on the sphere times time. These should be finite.
Models introduced by Myers et al. in \cite{Myers:2016wsu} produce a non-trivial expectation of a scalar operator of dimension $\Delta$ by having an interaction in the bulk that couples the dual scalar field to the curvature squared in the bulk. Such couplings can be argued to arise directly from string theory (the associated particle can decay to two gravitons, similar to dilaton couplings). The model has been solved in the flat brane case, with an expression that only converges
if $\Delta$ is bounded above by a number of order one \cite{Myers:2016wsu}.
  Grinberg and Maldacena \cite{Grinberg:2020fdj} argued that the computation admits an analytic continuation that gives rise to an expression whose dependence on $\Delta$ has the functional form
\begin{equation}
  \langle O\rangle\propto \frac{(d-2)(d-1)^2}{d }\frac{h(1-h)}{\sin(\pi h)}\frac{\Gamma (h)^2}{\Gamma (2 h)}
\end{equation}
where $h=\Delta/d$ is a normalized dimension of the operator, where $d$ is the dimension of the conformal field theory. In their paper they argued that the poles in $h$ arise from operator mixing. We want to understand this issue better.

In this paper we extract the same information by using the Frobenius method to solve the problems. In this case, the method itself does not require an analytic continuation and the answer is finite, except for the presence of the poles. The analysis readily produces generalized hypergeometric functions from the source that need to be expanded carefully around the horizon, where we extract a logarithmic divergence. Regularity at the horizon cancels the logarithm with the normalizable (at the boundary) solution of the unsourced equation, that has a similar singularity at the horizon.

The end result is similar to what occurs in conformal perturbation theory, where the gravity answer is automatically finite in calculations that usually require an analytic continuation in the conformal dimension of the operator  \cite{Berenstein:2014cia}. 

The presence of the poles that appear in the finite answer can be explained by the theory of how differential equations are solved around the boundary of AdS, given by a theorem of Fuchs. We will give a way to see this effect more explicitly by converting the radial direction into a fictitious time. The pole can be interpreted as arising from a forced harmonic oscillator driven at resonance, just like certain logarithmic behaviors can be interpreted as resonances in conformal perturbation theory \cite{Berenstein:2016avf}. With a little more care, one can check that the picture of operator mixing of \cite{Grinberg:2020fdj} can be understood explicitly.

Other problems, like the determination of response functions to the presence of a modulated scalar source at finite temperature  are also interesting (see for example \cite{Horowitz:2012ky,Horowitz:2012gs,Donos:2013eha, Hartnoll:2014cua}).  They are very hard to solve analytically, even in the linear regime. These response functions are usually solved numerically.  In the special case of two dimensional conformal field theories ($AdS_3$ black holes), the classic computation 
\cite{Son:2002sd} solved this linear problem. Recently, the retarded Green's function of $AdS_5$ black holes has also been computed \cite{Dodelson:2022yvn,Bhatta:2022wga} by converting the problem into solutions of the Heun equation. Some progress has also been made in the computation of quasinormal modes for these setups \cite{Bianchi:2021mft} (see also \cite{BarraganAmado:2018zpa}). The main insight is to relate the problem in gravity in $AdS_5$ to other problems in mathematical physics \cite{Aminov:2020yma}. In particular, these analytic computations require the ``connection formulae" for Heun functions expanded around different singularities of the equation For Kerr black holes it is found here \cite{Bonelli:2021uvf}. A more general study of some of the connection formulae for Heun functions can be found in \cite{Bonelli:2022ten}.

 In this paper we make progress on some of these 
questions by using a perturbative method for small momenta.
At first order the problem reduces to understanding singularities of more complicated sums of generalized hypergeometric functions. 

We use the same techniques to solve the one point function problem with a Weyl squared tensor source in global AdS black holes, rather than the flat Poincare black holes by using perturbation theory in the curvature of the sphere for large mass black holes. Here, the analysis of the singular behavior near the horizon is complicated by the fact that pieces of the perturbation term are more singular than in other examples. We show how to deal with these in detail and are able to obtain a first order correction.

The paper is organized as follows. In section \ref{sec:flatbr} we study the response function at zero momentum and the one point function sourced by the Weyl term using the method of Frobenius around infinity. In section \ref{sec:general} we study a slight generalization of this problem to arbitrary powers of the Weyl curvature. This calculation can be used to solve more generally the problem of response functions and one point functions  in a perturbative treatment that we perform. This includes the study of the spherical black hole. Additional issues arise in this case because the horizon gets shifted. In a perturbative treatment, we show how to deal with these to find a finite answer. We then conclude.

\section{The flat black brane case}\label{sec:flatbr}

The metric of the AdS flat black brane in $d+1$ dimensions is

\begin{equation}
    ds^2=\frac{L^2}{z^2}\left[-f(z) dt^2+ \frac{1}{f(z)}dz^2+d\vec{x}^2\right]
\end{equation}
where the function $f$ is given by
\begin{equation}\label{eq:fz}
    f(z)=1-\left(\frac{z}{z_{0}}\right)^{d}
\end{equation}
We want to study the static solution of a massive scalar field in this background, which is sourced by the square of the Weyl tensor. This coupling arises from the action 
\begin{equation}
   S=-\frac{1}{16 \pi G_{N}}\int \left[\frac{1}{2}(\nabla \phi)^2+\frac{1}{2} m^2 \phi^2+\alpha \phi W^2\right]
\end{equation}
as studied by Myers at al. and Grinberg and Maldacena. Our goal is to solve this problem using different techniques than what has been used so far. This will help understand in more detail why certain singularities can appear in the answer. The solution to this problem is straightforward from theorems in the study of linear differential equations. What is important is that the physics can be attributed to operator mixing with the background stress tensor.

We need the Weyl tensor squared as a  source. In these coordinates it is simple to write and given by 
\begin{equation}\label{Weyl}
    W^2=\frac{d(d-2)(d-1)^2}{L^4}\left(\frac{z}{z_{0}}\right)^{2d}
\end{equation}
Also, notice that when $z_0=\infty$ we get the metric of $AdS$ in the Poincar\'e patch. The region $z\simeq 0$ is the $AdS$ boundary. Whereas $z= z_0$ is the horizon: the locus where $g_{tt}$ vanishes with a single regular zero. The temperature can be related to $z_0$ as follows
\begin{equation}\label{scaling}
z_{0}^{-\Delta}=\left(\frac{4\pi T}{d}\right)^{\Delta}
\end{equation}

The expectation value of the stress tensor is proportional to the deviation away from the $AdS$ metric near $z\sim 0$. The subleading difference is captured by the decay of $g_{tt}$ near $z\sim 0$
\begin{equation}
    \langle T_{00}\rangle  \simeq \frac{1}{z_0^d}
\end{equation}
This metric is not in Fefferham-Graham coordinates, which would require that 
$g_{zz} = L^2/z^2$ for all $z$, but it is in a set of coordinates that make the analysis simpler to perform. In part this is because the determinant factor is simple 
\begin{equation}\label{determinant}
    \sqrt{-g}=\left(\frac{L}{z}\right)^{d+1}
\end{equation}
as it is only a power of $z$.

When we consider the action of the scalar field, we get the following equation of motion

\begin{equation}\label{EqOFmotion}
   \frac{1}{\sqrt{-g}}\partial_{\mu}\left(\sqrt{-g} g^{\mu \nu}\partial_{\nu}\phi\right)- m^2 \phi=\alpha W^2.
\end{equation}

We now want to consider a position and time independent solution of this equation, for various values of the mass $m^2$, or equivalently, using  the dimension of the associated operator $m^2L^2=\Delta (\Delta-d)$, 
the equation we need to solve is
\begin{equation}
   z^{d+1}\frac{d}{dz}\left(z^{-d+1}f(z)\frac{d\phi}{dz}\right)- \Delta (\Delta-d)\phi=\alpha L^2 W^2
\end{equation}
This equation depends explicitly on $d$, but we can eliminate any dimensional dependence after using the following change of variables 
\begin{equation}\label{eq:chofvars}
   \omega\equiv\left(\frac{z}{z_{0}}\right)^{d}, \quad h\equiv\frac{\Delta}{d}
\end{equation}
This way we obtain the following differential equation
\begin{equation}\label{ODE}
  \omega^2 \frac{d^2\phi}{d\omega^2 }- \omega \left[\omega^2\frac{d^2\phi}{d\omega^2 }+\omega\frac{d\phi}{d\omega}\right]+(1-h)h\phi=B\omega^2
\end{equation}
where the constant $B$ is given by
\begin{equation}\label{constant}
  B\equiv\frac{\alpha(d-2)(d-1)^2}{d L^2}.
\end{equation}

Notice that the term in square brackets is proportional to $\rho= \langle T_{00}\rangle $. That is, dimensional analysis in the variable $\omega$ is equivalent to counting powers of the expectation value $\rho$ (the energy density). In the coordinate $\omega$, there is a singularity of the differential equation at $\omega=0$ and another one at $\omega=1$ which is inherited from the coordinate singularity of the horizon. There is also a singularity at $\omega=\infty$. In a problem of linear differential equations, with three singularities in the complex plane at $\omega=0,  \omega=1, \omega=\infty$, the solution is closely related to hypergeometric functions. 

We will now solve the differential equation by the method of Frobenius. We need the homogeneous solution with regular boundary conditions at $z=0$ (the boundary of AdS) and a solution of the inhomogenous problem (also called the particular solution for the differential equation). We then combine both of these by requiring regularity at the horizon.

Let us start with the homogeneous solution (the one where the source is ignored). We write it as
\begin{equation}
 \phi_{H}= \omega^{\lambda}\sum^{\infty}_{n=0}a_{n}\omega^n
\end{equation}
and we normalize the solution so that $a_0=1$. The method of Frobenius will produce two results. First, it will solve for the value of $\lambda$ and then it gives a  recurrence relation for the $a_n$. 

The most singular term in the differential equation is the one of dimension $0$ in powers of $\omega$. The eigenvalue condition for $\lambda$ is given by
\begin{equation}
\lambda(\lambda-1) - h(h-1) =0,
\end{equation}
which has two clear solutions: $\lambda = h$ and $\lambda = 1-h$. We assume for simplicity that $\lambda=h>0$. When $\lambda<0$ the solution diverges at $\omega=0$ and is considered unphysical from the point of view of the problem we are considering (more generally a source has been added for the field $\phi$ on the boundary of $AdS$).

We also obtain the following recurrence relation
\begin{equation}\label{eq:rec}
    \frac{a_{n+1}}{a_{n}}=\frac{(n+h)^2}{(n+2h)(n+1)}.
\end{equation}

We recognize the general coefficient $a_{n}$ as
\begin{equation}
    a_{n}=\frac{1}{n!}\frac{(h)_{n}(h)_{n}}{(2h)_{n}}.
\end{equation}
Here we have introduced the Pochhammer symbol defined as
\begin{equation}
  (y)_{n}\equiv\frac{\Gamma(y+n)}{\Gamma(y)}=y(y+1)...(y+n-1).
\end{equation}
The solution of the homogeneous term is 
\begin{equation}\label{Hserie}
    \phi_{H1}=  \omega^h \sum^{\infty}_{n=0}\frac{1}{n!}\frac{(h)_{n}(h)_{n}}{(2h)_{n}}\omega^n=\omega^{h} {}_{2}F_{1}(h,h,2h,\omega),
\end{equation}
that we immediately recognize as a $\ _2F_1$ hypergeometric function.
Since the horizon is located at $\omega=1$, it is convenient to understand the large $n$ limit of the terms in the series, to get at the structure of the possible divergence at $\omega=1$. We find that since
\begin{equation}
    \lim_{n\to \infty} \frac{\Gamma(n+h)^2}{\Gamma(n)\Gamma(n+2h)}=1
\end{equation}
the series for large $n$ looks like
\begin{equation}
 \phi_{H_1}\sim \frac{\Gamma(2h)}{\Gamma(h)^2} \sum_{n}^\infty \frac1{n} \omega^n \sim -\frac{\Gamma(2h)}{\Gamma(h)^2} \log(1-w)\label{eq:logdiv1}
\end{equation}
so that it diverges logarithmically. The  series expansion around $\omega=1$ that uses the full analytic properties of the hypergeometric function is
\begin{eqnarray}
\phi_{H_1}&\approx&  -\frac{ \Gamma (2 h) }{\Gamma (h)^2}(2 \psi ^{(0)}(h)+\log (1-\omega)+2 \gamma )\nonumber\\
&&+\frac{(h-1) h  \Gamma (2 h)}{\Gamma (h)^2} (2 \psi ^{(0)}(h)+\log (1-\omega)+2 \gamma -2)(\omega-1)+O\left((\omega-1)^2\right).
\end{eqnarray}
Here $\gamma$ is the Euler-Mascheroni constant and $\psi ^{(0)}(h)$ is the first Polygamma Function (Digamma Function). These extra finite pieces are not relevant for the computation of the one point function of the boundary.

One can similarly consider the $h\to 1-h$ homogeneous solution that diverges at the boundary.
In that case the logarithmic term diverging at the horizon will be given by
\begin{equation}\phi_{H_2}\sim -\frac{\Gamma(2-2h)}{\Gamma(1-h)^2 }\log(1-w).
\end{equation}
The most general solution of the homogeneous differential equation part of (\ref{ODE}) is given by
\begin{equation}
  \phi_{T}=c_{1}\phi_{H1}+c_2\phi_{H2}
\end{equation}
We
require that the linear combination of $\phi_{H_1}$ and $\phi_{H_2}$ is regular at $\omega=1$ (the logarithmic divergence cancels between them). This is how regularity at the horizon is imposed. Basically, the divergent piece of the stress tensor for the scalar field $\phi$ that is proportional $(\partial_w\phi)^2 \sim \frac 1 {(1-w)^2}$ is absent.

The response function for a source at the boundary,  at finite temperature will be proportional to $-c1/c2$. 
In this case
\begin{equation}
\langle\delta \phi_{H_1}\rangle_{\delta \phi_{H_2}} = -\frac{\Gamma(h)^2}{\Gamma(1-h)^2}\frac{\Gamma(2-2h)}{\Gamma(2h)}.   
\end{equation}

The temperature dependence (dependence on $\rho$, or equivalently $z_0$) is fixed by dimensional analysis and a correction factor from normalization  $d(2h-1)= 2\Delta -d$ (see the discussion in \cite{Marolf:2004fy,Berenstein:2014cia}). 
This way
\begin{equation}
    \left\langle \frac{\delta \phi}{\delta J}\right \rangle= -\frac{\Gamma(h)^2}{\Gamma(1-h)^2}\frac{\Gamma(2-2h)}{\Gamma(2h)} (2\Delta -d) z_0^{d-2 \Delta}
\end{equation}
Notice that the answer can be divergent when $h$ is a half-integer. Then $\Gamma(2-2h)$ is divergent while the  terms in the denominator are regular.
 The divergence appears already from the Pochhammer symbol in equation \eqref{Hserie}  where we find a denominator $n+2(1-h)=0$.
The power of $\omega$ appearing for this term is $\omega^{(1-h)+n+1}= \omega^h$ and it 
coincides with the positive root for $\lambda$, namely $h$.
This has nothing to do with the horizon. Instead, it indicates that there is a na\"\i ve failure of the Frobenius method.
In that case the two roots $h, 1-h$ differ by an integer. 
This is resolved by having a series with an additional $\phi_{H_1}\log(w)$ starting at this term. This case is a special example of  Fuchs' theorem.
 For $h$ integer, the denominator will have a double pole at $n=h$, which cancels the 
 pole in the numerator and makes the answer vanish. What happens is that the recurrence relation \eqref{eq:rec} terminates when $n+1-h=0$, before we hit the singularity $n+2-2h=0$.
 In that case, the solution with the boundary source is already regular at the horizon and one can not turn on a non-trivial response function.
The normalization of the field in AdS is more subtle, but one can compare the field theory two point function and the AdS bulk two point function to do that \cite{Berenstein:1998ij,Berenstein:2014cia}, resulting in an extra factor of
\begin{equation}
    {\cal N}=\frac{\Gamma(\Delta)}{2\pi^{d/2}\Gamma(\Delta-d/2+1)}
\end{equation}
that has no additional poles in $h$, as $\Delta > 1-d/2 $. From our point of view, this is a global rescaling factor that has nothing to do with the details of the differential equation, so we can drop it.

We now want to consider the bulk source, rather than the boundary source. The techniques for solving the problem are essentially identical, except that we will not get a $\ \!_2F_1$, but more complicated functions. 

\subsection{Particular solution $\phi_{P}$}

Now, we focus on obtaining the particular solution $\phi_{P}$ of the differential equation (\ref{ODE}) with the squared Weyl tensor as a source. We look for a series solution
\begin{equation}
    \sum^{\infty}_{n=0}a_{n} \omega^{n+\lambda} [(1-h)h+ (n+\lambda)(n+\lambda-1)]-\sum^{\infty}_{n=0}a_{n} \omega^{n+\lambda+1} (n+\lambda)^2=B \omega^2.
\end{equation}
To match the $\omega^2$ dependence of $B\omega^2$, we need to start with $\lambda=2$.
The coefficient $a_0$ is given by
\begin{equation}\label{eq:a0}
    a_{0} = \frac{B}{(2-h)(1+h)}.
\end{equation}
The recurrence relation is
\begin{equation}
      a_{n+1} =\frac{(n+2)^2}{(n+2+h)(n+3-h)} a_{n}
\end{equation}
and it is a ratio of polynomials of $n$. The answer is written in terms of a $\ \!_3F_2$ series, as follows
\begin{equation}\label{eq:recpart}
    a_{n}=\frac{(2)_{n}(2)_{n} }{(3-h)_{n}(h+2)_{n}}a_0 =\frac{1}{n!}\frac{(1)_{n}(2)_{n}(2)_{n} }{(3-h)_{n}(h+2)_{n}}a_{0}
\end{equation}
where we use $(1)_n= n!$. The solution is explicitly written as
\begin{equation}\label{PhiPG}
    \phi_{P}= \frac{B \omega^2}{(2-h) (h+1)}{}_{3}F_{2}([1,2,2],[3-h,h+2],\omega)
\end{equation}
Again, we can look at the asymptotic behavior of the coefficients by using the limit
\begin{equation}
    \lim_{n\to \infty} \frac{\Gamma(n+1) \Gamma(n+2) \Gamma(n+2) }{\Gamma(n)\Gamma(n+3-h) \Gamma(n+h+2) }=1
\end{equation}
to find that
\begin{eqnarray}
    \phi_P &\sim& \frac{B \omega^2}{(2-h) (h+1)}\left. \left[\frac{ \Gamma(n+2) \Gamma(n+2) }{\Gamma(n+3-h)\Gamma(n+h+2)}\right]^{-1}\right|_{n=0}\sum \frac{w^n}{n}\nonumber\\
    &\sim&- \frac{B \omega^2}{(2-h) (h+1)} \left[\frac{ \Gamma(2) \Gamma(2) }{\Gamma(3-h)\Gamma(2+h)}\right]^{-1}\log(1-\omega)\nonumber\\
    &\sim& -B\Gamma(2-h)\Gamma(1+h)\log(1-\omega) \label{eq:phipsing}]
\end{eqnarray}
We want to re-express this limit in a more illuminating way, so we recall the following identity for Gamma functions
\begin{equation}
    \Gamma (1-z) \Gamma (z)=\frac{\pi}{\sin(\pi z)}
\end{equation}
Therefore, we obtain the following expression
\begin{equation}\label{limit2}
    \lim_{\omega \to 1-\epsilon}\phi_{P}=-B\frac{\pi(1 - h) h}{\sin(\pi h)}\log (\epsilon)
\end{equation}

The most general solution of the differential equation (\ref{ODE}) is given by
\begin{equation}
  \phi_{T}=c_{1}\phi_{H1}+\phi_{P}
\end{equation}
The thermal one-point function is proportional to $c_{1}$, and we can determine this constant by demanding regularity for $\phi_{T}$ at the horizon. So, taking the $\omega\rightarrow 1$ limit we obtain that the divergent pieces must cancel
\begin{equation}\label{LogLimit}
   \lim_{\omega \to 1-\epsilon}\phi_{T}|_{sing}=-c_{1}\frac{ \Gamma (2 h) }{\Gamma (h)^2}\log (\epsilon)-B\frac{\pi(1 - h) h}{\sin(\pi h)}\log (\epsilon)
\end{equation}
 Demanding that this limit is regular (that is, the logarithmic divergences cancel), we obtain that
\begin{equation}
   c_{1}= -B\pi\frac{h(1-h)}{\sin(\pi h)}\frac{\Gamma (h)^2}{\Gamma (2 h)}
\end{equation}
Finally, we take into account the physical scale (\ref{scaling}) to obtain the following  thermal one-point function  

\begin{equation}
  \left<O\right>\propto-\left(\frac{4\pi T}{d}\right)^{\Delta}\frac{\alpha\pi(d-2)(d-1)^2}{d L^2}\frac{h(1-h)}{\sin(\pi h)}\frac{\Gamma (h)^2}{\Gamma (2 h)}
\end{equation}

Except for some overall factors associated with the normalization of the massive scalar action and the normalization of the one-point function, we have replicated the result of Maldacena-Grinberg  \cite{Grinberg:2020fdj}, which generalized results from Myers et al. \cite{Myers:2016wsu} that were computed using the Greens function method by analytic continuation. Here we see that 
in this treatment the answer does not need additional regularization or analytic continuation.

Notice that there can be singularities in the answer: poles when $h$ is integer appearing from $\sin(\pi h)$. These  can be traced to the $\Gamma(2-h)$ appearing in \eqref{eq:phipsing}, or equivalently, to the $(3-h)_n$ symbol appearing in \eqref{eq:recpart}. These indicate a pole at some finite value of $n$: again, these are indicative of the naive failure of the Frobenius method as an expansion around $\omega=0$
when we are studying the problem at finite order. These have nothing to do with the singular behavior at the horizon. The fix requires a logarithmic correction proportional to $\log(\omega)$ times $\phi_{H_1}$ to the answer. These occur when $h$ is a positive integer greater than or equal to 3, and render the one point function ambiguous: $\log(\omega)$ can be replaced by $\log(\omega \tilde \Lambda)$ for any $\tilde \Lambda$, thus shifting the value of  $c_1$ by a finite amount.

If we think of $\log(\tilde \Lambda)$ as a cutoff in quantum field theory (it determines a surface in the $UV$  by $\log(\omega \tilde \Lambda)=0$), we see that the answer bears close similarity to logarithmic divergences in quantum field theory. Where can these arise from?
The answer suggested in \cite{Grinberg:2020fdj} is that they arise from mixing between a power of the stress tensor $:T^s:$ and ${\cal O}$. We would like to explain this intuition in more detail.

The idea is the following. Consider a new coordinate $\sigma$ defined by  $\omega= \exp(i \sigma)$. The differential equation \eqref{ODE} becomes
\begin{equation}
    -\partial_\sigma^2\phi + i \partial_\sigma\phi  +(1-h)h \phi= B \exp(2i \sigma) +\exp(i\sigma)\left[ -\partial_\sigma^2\right]\phi.
\end{equation}
The $\sigma$ coordinate is behaving like a time, rather than a position, because we added the factor of $i$. This coordinate lets us interpret the radial direction $\omega$ as a flow in a fictitious time. 

The left hand side is translation invariant in $\sigma$. The right hand side has two pieces: a source at constant frequency $\exp(2i \sigma)$, and a field dependent source that adds one unit to the frequency. This second term is the one that is proportional to the energy density $\rho$. The particular solution is
\begin{equation}
\sum a_n \exp(i (n+2) \sigma)
\end{equation}
We see that the singularity arises from a resonance condition, when 
$(n+2)^2-(n+2) +h(1-h)= (n+2-h)(n+1+h)= 0$. This is the condition to be on-shell for the $\sigma$ independent part of the differential equation: if we are driving at resonance, 
the  solution must grow linearly in the time $\sigma$ on top of the oscillating piece.
This is the $\log(\omega)$ we had before. The precise solution depends on ``initial conditions". Notice that this occurs on a term that behaves as $\rho^{n+2}$, since  the
Weyl tensor squared also counts as two powers of $\rho$. The resonance is between a quantity proportional to $\langle :T_{00}:^{n+2}\rangle$ and the operator dual to the field $\phi$: they both have the same dimension when $h$ is a positive integer greater than or equal to 2. The normal ordering here indicates that we have the naive dimension of the composite operator $T_{00}^{n+2}$. Basically, the terms in the series expansion of the particular solution, with the source $W^2$ seeding the solution are oscillating at integer frequencies. When $h$ is not an integer there is no resonance condition and in principle there is no operator mixing ambiguity that needs to be considered. This is very similar to how logarithmic divergences from 
mixing appear in other settings \cite{Berenstein:2016avf}

\section{Some simple generalizations}\label{sec:general}

In our next step, we consider the following action for a massive scalar field
\begin{equation}
   S=-\frac{1}{16 \pi G_{N}}\int \left[\frac{1}{2}(\nabla \phi)^2+\frac{1}{2} m^2 \phi^2+\alpha \phi W^{2\beta}\right].
\end{equation}
Basically, we are writing a solution with a source that has a different power of $z$ than the one that arises from the Weyl squared tensor $W^2$. Many other similar problems can be analyzed if we have a better understanding of solutions to the modified source as a more general power of $z$. In the $w$ variables, the differential equation is the following
\begin{equation}
 (1-\omega)\omega^2 \frac{d^2\phi}{d\omega^2}-\omega^2\frac{d\phi}{d\omega}+(1-h)h\phi=K \omega^{2\beta} \label{eq:modbeta}, 
\end{equation}
where we have defined the constant $K$ as follows
\begin{equation}
    K\equiv \left(\frac{\alpha R^2}{d^2}\right)\left(\frac{d(d-2)(d-1)^2}{R^4}\right)^{\beta}
\end{equation}
Again, at this stage, what we are really doing is solving for an arbitrary power of $\omega$ in the right hand side of equation \eqref{eq:modbeta}. The inhomogenous solution will start with a  power of $\omega^{2\beta}$, rather than $\omega^2$. The result for the first term is very similar to \eqref{eq:a0}, giving
\begin{equation}
    a_{0} = \frac{K}{(h+{2\beta} -1)({2\beta}-h )}.
\end{equation}
The full result, after following the method of Frobenius is:
\begin{equation}
    \phi_{P_{\beta}}= \frac{K \omega^{2\beta}}{(2\beta-h) (h+2\beta-1)}\sum^{\infty}_{n=0}\frac{1}{n!}\frac{(1)_{n}(2\beta)_{n}(2\beta)_{n} }{(1+2\beta-h)_{n}(h+2\beta)_{n}}\omega^n.\label{eq:arbsource}
\end{equation}
This is again, a $\ \!_3F_2$ function with different parameters:
\begin{equation}
    \phi_{P_{\beta}}= \frac{K \omega^{2\beta }}{(2\beta-h) (h+2\beta-1)}{}_{3}F_{2}([1,2\beta,2\beta],[1+2\beta-h,h+2\beta],\omega).
\end{equation}
Using the same methods as before, we get that  the near horizon limit logarithmic behavior is given by 
\begin{equation}\label{eq:loglimit3f2}
    \lim_{\omega \to 1-\epsilon}\phi_{P_{\beta}}=- K\frac{  \Gamma (2 \beta -h) \Gamma (h+2 \beta -1)}{\Gamma (2 \beta)^2}\log (\epsilon)
\end{equation}
It is then straightforward to find the one point function by requiring that the logarithmic divergence of the inhomogeneous solution cancels with the homogeneous piece as well. 

\subsection{Response function to a weakly modulated source}

We now consider a solution that apart from radial variable $z$ also has dependence on the spatial variables $x^i$, so that $\phi=\phi(z,x)$. We notice that we still consider a static solution, but allow for modulation (slow variation in $x$). The Equation of motion in this case is 
\begin{equation}
   \frac{1}{\sqrt{-g}}\partial_{z}\left(\sqrt{-g} g^{z z}\partial_{z}\phi\right)+g^{ii}\partial^2_i \phi- m^2 \phi=\alpha W^2.
\end{equation}

By doing a Fourier transformation in the variables $x^i$, and using a spatial profile $\exp(i \vec k \cdot \vec x)$, we can decompose into momentum modes. The $x$ dependence becomes trivial and we get a slight modification of the equation
\begin{equation}
   \omega^2 \frac{d}{d\omega}\left((1-\omega)\frac{d\tilde{\phi}}{d\omega}\right)+(1-h)h\tilde{\phi}-q^2\omega^{2/d}\tilde{\phi}=B_W \omega^2\delta^{d-1}(k)
\end{equation}
where $q$ is the momentum variable in dimensionless units $q\equiv\frac{k}{4\pi T}$. The idea is to expand perturbatively in $q$, in the presence of a weakly modulated source. We have also used the radial variable $\omega$ defined in (\ref{eq:chofvars}). Since we are interested in the case $k\neq0$, we can safely ignore the Weyl source. Now, we assume the regime $q<<1$ in order to solve the previous equation perturbatively $\tilde{\phi}_{Hi}(u,q)=\phi_{Hi}^{(0)}(u)+\delta \tilde{\phi}_{Hi}(u,q)$ where we notice that $\tilde{\phi}_{Hi}^{(0)}(u)=\phi_{Hi}^{(0)}(u)$. Thus, the differential equation becomes
\begin{equation}\label{Hom spatial case}
 \omega^2 \frac{d}{d\omega}\left(\left(1-\omega\right)\frac{d(\delta \tilde{\phi}_{Hi})}{d\omega}\right)+(1-h)h (\delta \tilde{\phi}_{Hi})=q^2\omega^{2/d}\phi_{Hi}^{(0)} 
\end{equation}

We can write the homogeneous solutions given in \eqref{Hserie} in terms of series as follows, for both $h$ and $1-h$ (so that we have the two solutions without a source in the bulk)

\begin{equation}\label{an}
    \phi_{H1}^{(0)}=\omega ^h\sum a_n \omega ^n, \quad a_n=\frac{1}{n!}\frac{(h)_n (h)_n}{(2h)_n},
\end{equation}
\begin{equation}\label{bn}
    \phi_{H2}^{(0)}=\omega^{1-h}\sum b_n \omega^n, \quad b_n=\frac{1}{n!}\frac{(1-h)_n (1-h)_n}{(2-2h)_n}.
\end{equation}

From here, we notice that the equation \eqref{Hom spatial case} becomes exactly the same differential equation given by \eqref{eq:modbeta} (on a term by term basis)  by the identification of parameters:  $\beta_{H1}=n+h+2/d$, $K_{\beta_{H1}}=q^2 a_n$ and
$\beta_{H2}=n+1-h+2/d$, $K_{\beta_{H2}}=q^2 b_n$.
We can just write sums of the corresponding solutions \eqref{eq:arbsource} to get the answer at this order.
Now, we use the near horizon limit given in (\ref{eq:loglimit3f2}) term by term to get the logarithmic divergence

\begin{equation}
    \lim_{\omega \to 1-\epsilon}\delta \tilde{\phi}_{H1}(u,q)=- q^2\sum a_n\frac{\Gamma (n+2/d) \Gamma (n+2h-1+2/d)}{\Gamma (n+h+2/d)^2}\log (\epsilon)\equiv - q^2\sum A_n\log (\epsilon)
\end{equation}

\begin{equation*}
    \lim_{u \to 1-\epsilon}\delta \tilde{\phi}_{H2}(u,q)=- q^2\sum b_n\frac{\Gamma (n+2/d) \Gamma (n-2h+1+2/d)}{\Gamma (n+1-h+2/d)^2}\log (\epsilon)\equiv - q^2\sum B_n\log (\epsilon)
\end{equation*}

We have ignored in both limits an infinite sum of finite terms that could potentially diverge. This can be analyzed more carefully from \eqref{Hom spatial case}. The idea is that the right hand side is smooth for $\omega<1$ and decays rapidly for $\omega \sim 0$ (faster than $\phi^{(0)}_{H_i}$). The most dangerous part of the analysis is that $\phi^{(0)}_{H_i}$
diverges at $\omega\to 1$ logarithmically, so $\delta\phi$ might be more singular than we originally thought. It can be checked by expanding directly around $\omega=1$ that if $\phi_{H_i}^{(0)}$ behaves logarithmically, then the terms of $\delta \phi$ near $\omega\to 1$ grow at most logarithmically.
This will come from the $h(1-h)$ term. Basically, the second order differential equation is elliptic in the region $\omega\in (0,1)$, so the solution will not be more singular than the singularity of the source.

The derivative terms in fact  kill a logarithmic divergence of $\delta\phi$. To get a log on the left hand side from the second derivative would require a term $(1-\omega) \log(1-\omega)$, which has a zero limit when $\omega\rightarrow 1$ and is less singular than the right hand side. The term with $h(1-h)$ would then be the one that
dominates the first term in the expansion around $\omega\sim1$.
We conclude that in practical terms we can ignore those finite terms safely. Thus, the limit of the total homogeneous solution is 
\begin{equation}
   \lim_{\omega \to 1-\epsilon} \tilde{\phi}(u,k)=\left[-c_1\left(\frac{\Gamma(2h)}{\Gamma(h)^2}+q^2\sum A_n\right)-c_2\left(\frac{\Gamma(2-2h)}{\Gamma(1-h)^2}+ q^2\sum B_n\right)\right]\log (\epsilon)
\end{equation}
Demanding the regularity condition at the horizon, we obtain the following response function ( which is given by $-\frac{c_1}{c_2}$ times certain normalization factors independent of $k$ )
\begin{equation}
     -\frac{c_1}{c_2}=\frac{\Gamma(h)^2}{\Gamma(2h)}\frac{\Gamma(2-2h)}{\Gamma(1-h)^2}\left[1+q^2\left(\frac{\Gamma(1-h)^2}{\Gamma(2-2h)}\sum B_n-\frac{\Gamma(h)^2}{\Gamma(2h)}\sum A_n\right)\right].
\end{equation}
These sums can be re-expressed in terms of regularized hypergeometric functions of type $_{4}\tilde{F}_3$ given by
\begin{multline}\label{A}
   \mathcal{A}(h,d)\equiv \frac{\Gamma(h)^2}{\Gamma(2h)}\sum A_n=\Gamma(2/d)\Gamma(h)^2\Gamma(2/d+2h-1) \\ _{4}\tilde{F}_3[(h,h,2/d,2/d+2h-1),(2h,h+2/d,h+2/d);1],
\end{multline}
and 
\begin{multline}\label{B}
    \mathcal{B}(h,d)\equiv  \frac{\Gamma(1-h)^2}{\Gamma(2-2h)}\sum B_n=\Gamma(2/d)\Gamma(1-h)^2 \Gamma(2/d+1-2h)\\ _{4}\tilde{F}_3[(1-h,1-h,2/d,2/d-2h+1),(2-2h,1-h+2/d,1-h+2/d);1].
\end{multline}

By analyzing the parameters of these hypergeometric functions, one can infer that these functions are regular at $1$. Therefore, by removing the logarithm divergence we are assured to get a finite result. Finally, the response function is given by
\begin{equation}\label{Response Function}
    \left\langle \frac{\delta \phi}{\delta J}\right \rangle=-(2\Delta -d) z_0^{d-2 \Delta}\frac{\Gamma(h)^2}{\Gamma(2h)}\frac{\Gamma(2-2h)}{\Gamma(1-h)^2}\left[1+q^2\left(\mathcal{B}(h,d)-\mathcal{A}(h,d)\right)\right]
\end{equation}
We also notice that in this computation, we didn't use the Weyl source which is defined only for $d\geq3$. Therefore, we can evaluate the answer at $d=2$ to obtain that
\begin{equation*}
     \left\langle \frac{\delta \phi}{\delta J}\right \rangle^{d=2}\propto\frac{\Gamma(h)^2}{\Gamma(2h)}\frac{\Gamma(2-2h)}{\Gamma(1-h)^2}\left[1+q^2\left(\psi^{(1)}(1-h)-\psi^{(1)}(h)\right)\right]
\end{equation*}
On the other hand, for the planar BTZ black hole, the retarded Green function (See for instance \cite{Son:2002sd, Blake:2019otz}) is given by
\begin{equation}\label{eq:polesBTZ}
    G_r(q,\nu)\propto \frac{\Gamma(2-2h)}{\Gamma(2h)} \frac{\Gamma(h+i(\nu-q))\Gamma(h+i(\nu+q))}{\Gamma(1-h+i(\nu-q))\Gamma(1-h+i(\nu+q))}
\end{equation}
where $\nu\equiv\frac{\omega}{4\pi T}$. Expanding for small momenta, we obtain 
\begin{equation*}
     G_r(q,\nu=0)\propto \frac{\Gamma(2-2h)\Gamma(h)^2}{\Gamma(2h)\Gamma(1-h)^2}\left[1+q^2\left(\psi^{(1)}(1-h)-\psi^{(1)}(h)\right)\right] +O(q^3)
\end{equation*}
Therefore, our method reproduces at first order in $q^2$ the pure spatial part of the retarded Green function of the planar BTZ black hole. This is a consistency check that our result in (\ref{Response Function}) is correct and provides the first order term in any dimension $d$ of the purely spatial response function.

One can ask, from our point of view, what makes the case $d=2$ simple? When $d=2$, the exponent of $\omega$ on the $q^2$ term is $\gamma= 2/d=1$ produces 
an additional term in $\exp(i\sigma)$ at the same frequency than the term that arises from the metric. In that sense, if one does not use a perturbation approach, one finds instead a modified quadratic numerator in the recurrence
\eqref{eq:rec} given by
\begin{equation}
    \frac{a_{n+1}}{a_n}= \frac{(h+n)^2 +q^2}{(n+2h)(n+1)}
\end{equation}
whose numerator can be easily factorized into $(h+n+iq)(h+n-iq)$. 
The solution of the recursion is
\begin{equation}
    a_n= \frac{ (h+i q)_n(h-iq)_n}{n! (2h)_n}
\end{equation}
and the answer is in terms of hypergeometric functions, where the parameters are modified by $q$
\begin{equation}
    \phi_h \sim \omega^h \ {}_2F_1(h+iq, h-iq, 2h,\omega )
\end{equation}
It is easy to check that all the poles in $q$ in the numerator and denominator at finite $q$ appearing in equation \eqref{eq:polesBTZ} are captured this way: the logarithmic behavior at $\omega\to 1$ cancels in the same way as when we work with $q=0$. In principle, one should be able to argue that the dependence on the frequency also comes automatically. The linear combinations $\nu \pm q = const$ are related to traveling at the speed of light, as $\partial_q \nu=\pm 1$, splitting into products of left and right moving modes. Moreover, the poles in the numerator (quasinormal modes) should occur for a fixed sign of the imaginary part of $\nu$,  $\Im m(\nu)$ at real momentum $q$. If one can argue that the left and right movers must be split this way, the full result should immediately follow from just the spatial modulation dependence.

\subsection{One point function in global AdS black holes in the large mass limit.}

We are going to consider again a massive scalar field but this time in a $AdS_{d+1}$ spherical black hole background rather than a black brane.
We do not know how to make the full calculation exactly. What we can do is a perturbative expansion in inverse powers of the mass (expanding around large black holes). In this case,  in the equation of motion we are going to consider the horizon curvature as a small parameter so as to be able to use perturbation theory around the black brane solution. 

The $AdS_{d+1}$ spherical black hole metric is
\begin{equation}
    ds^2=-H(r) dt^2+ \frac{1}{H(r)}dr^2+r^2 d\Omega^2,
\end{equation}
where $H(r)$ is defined as follows
\begin{equation}
    H(r)=1+\frac{r^2}{L^2}-\frac{2M}{r^{d-2}}.
\end{equation}
These coordinates are convenient as the determinant of the metric is 
\begin{equation}\label{DET}
    \sqrt{-g} \propto r^{d-1}
\end{equation}
and the Weyl tensor squared is given by a simple expression
\begin{equation}
    W^2=\frac{d(d-2)(d-1)^2}{r^{2d}}\left(2M\right)^{2}
\end{equation}
If we compare with the flat black brane, we would use instead
$H(r)= \frac{r^2}{L^2}-\frac{2M}{r^{d-2}}$, and the boundary of spacetime is located at $r\to \infty$. The natural $z$ coordinate that we used before is
simply $z= L^2/r$. In this way we find that 
\begin{equation}
    H(z) = L^2\left( \frac{1}{z^2}+ \frac{1}{L^2}- \frac 1{z^2}\left[\frac{z}{z_0}\right]^d\right)
\end{equation}
where $z_0$ is determined by looking at the flat brane factor without the curvature correction. Namely, we first define $r_0^d= 2 M L^2$. In this sense, $r_0$ is a notion of the radius of the black hole, and then we take $z_0\to L^2/r_0$.
This  is equivalent to having an adjusted value of $f(z)$ in equation \eqref{eq:fz}, given by
\begin{equation}
    f(z) = 1 -\left(\frac{z}{z_0}\right)^d + \frac{z^2}{L^2}  
\end{equation}
and the idea of a large mass black hole is that the additional term is
very small still when $z\sim z_0$, that is, we take $z_0/L \ll 1$ so the horizon is very close to the boundary. Notice that large $r_0$ is equivalent to large $M$, and that large $r_0$ becomes a small 
$z_0$ after the inversion.
We can now introduce the dimensionless 
$\omega = (z/z_0)^d$ coordinate that we 
used before. In this coordinate system we have that $z^2= z_0^2 \omega^{2/d}$.
After some algebra, the differential equation becomes
\begin{equation}\label{ODEspherical}
 \omega^2\left(s^2 \omega^{\gamma}+1-\omega\right) \frac{d^2\phi}{d\omega^2}+\omega\left(\gamma s^2  \omega^{\gamma}-\omega \right)\frac{d\phi}{d\omega}+(1-h)h\phi=B \omega^2
\end{equation}
where
\begin{equation}\label{s^2}
   \gamma\equiv\frac{2}{d} , \quad s^2\equiv\frac{L^2}{r_0^2}= \frac{L^2}{(2ML^2)^\gamma}.
\end{equation}
We now treat $s$ as a small parameter in which we can do perturbation theory.
We notice that if we take $s\to 0$ we recover our previous differential equation with the source, so $s$ can indeed be treated perturbatively.

It is important to understand at this point why the general problem is going to be hard to solve:
the equation \eqref{ODEspherical} has more singularities in the complex plane than the corresponding planar problem. These occur at $\omega=0, \infty$ and at the non trivial roots of $(s^2 \omega^{\gamma}+1-\omega)=0$.
There is one that is perturbatively close to $\omega=1$, but there are new singularities.
In practice this means that the problem can not be handled with hypergeometric functions. An exact solution becomes very hard 
in that case. This is why we will attempt to do a perturbative expansion in $s^2$. This is actually a non-trivial expansion.

\subsubsection{Perturbative computation for the homogeneous solution}

We write down the total homogeneous solution $\phi_{H}^{Tot}$ as follows

\begin{equation}\label{phiTotal}
   \phi_{H}^{Tot}=\phi_{H}+\delta \phi_{H},
\end{equation}
where $\phi_{H}$ is the homogeneous solution of the black brane that we already computed and is given by
\begin{equation*}
    \phi_{H}= \omega^{h} {}_{2}F_{1}(h,h,2h,\omega)= \omega^h \sum^{\infty}_{n=0}a_{n}\omega^n, \quad a_{n}=\frac{1}{n!}\frac{(h)_{n}(h)_{n}}{(2h)_{n}},
\end{equation*}
and where $\delta \phi$ is a first order correction proportional to $s^2$.
We plug the solution we already know, equation \eqref{Hserie} into the differential equation (\ref{ODEspherical}) to obtain 
\begin{equation}\label{ODEPH}
 \omega^2(1-\omega)\delta \phi_{H}''-\omega^2\delta \phi_{H}'+h(1-h)\delta \phi_{H}=-s^2 \omega^{\gamma}(\gamma\omega \phi_{H}'+\omega^2 \phi_{H}'').
\end{equation}

In this last expression, we have dropped terms of higher order $s^2 \delta \phi_{H}'$ and $s^2 \delta \phi_{H}''$, and used the homogeneous differential equation for $\phi_{H}$. Let us use the information we have about how $\phi_H(\omega)$ behaves as $\omega\to 1$. It depends on $\omega$ as $\phi_H \sim C \omega^h  \log(1-\omega)+ \omega^h C'+ \omega^h O(1-\omega)$. The second derivative of $\phi_H$
gives a double pole in $(1-\omega)$ on the right hand side. This needs to be cancelled by the most singular term in the differential equation for $\delta \phi$. This requires a divergence of $\delta \phi_H$ that behaves as $\Theta (1-\omega)^{-1}$, whose coefficient is uniquely determined by $C$. Since $C$ has already been computed, this most singular term is simple. 

At the next order, we get a single pole in $1-\omega$ on the right hand side. This extra pole can arise from a $\log(1-\omega)^2$ in $\delta\phi_H$ (a double logarihtmic singularity). The rest of the terms on the right hand side are at most log divergent and we have already seen that these are not a problem in the previous example.  A similar structure arises from the inhomogeneous solution and since the $\phi_H+\phi_P$ is regular at $\omega=1$.

Now, we compute the following quantities
\begin{eqnarray}
    \omega \phi_{H}'&=& \sum^{\infty}_{n=0}a_{n} (n+h)\omega^{n+h}\\
    \omega^2 \phi_{H}''&=& \sum^{\infty}_{n=0}a_{n} (n+h)(n+h-1)\omega^{n+h}
\end{eqnarray}
as power series in $\omega$. Each term essentially corresponds to a source with a different coefficient $K$ and a different $\beta$ in \eqref{eq:modbeta}, so we can use those results term by term, finding that
\begin{equation}
 -s^2 \omega^{\gamma}(\gamma\omega \phi_{H}'+\omega^2 \phi_{H}'')=-s^2 \sum^{\infty}_{n=0}a_{n} (n+h)(n+h+\gamma-1)\omega^{n+h+\gamma}
\end{equation}
To simplify this expression, we define $\bar{A}_{n}$ as 
\begin{equation}\label{An}
  \bar{A}_{n} \equiv  \frac{1}{n!}\frac{(h)_{n}(h)_{n}}{(2h)_{n}}(n+h)(n+h+\gamma-1)
\end{equation}
so that the problem to solve is the following differential equation
\begin{equation}
 \omega^2(1-\omega)\delta \phi_{H}''-\omega^2\delta \phi_{H}'+h(1-h)\delta \phi_{H}=-s^2 \sum^{\infty}_{n=0}\bar{A}_{n} \omega^{n+h+\gamma}.
\end{equation}
Notice that the $\bar A_n$ coefficients are more divergent when $n\to \infty$ than the ones for $a_n$. This is the source of the different singular behavior as $\omega\to 1$.

We now  need to sum over the solutions we already know to find that the modified homogeneous solution is
\begin{equation}
  \delta \phi_{H}=-s^2 \sum^{\infty}_{n=0}\frac{\bar{A}_{n} \omega^{n+h+\gamma}}{(n+\gamma)(n+\gamma+2h-1)} {}_{3}F_{2}([1,n+h+\gamma,n+h+\gamma],[1+n+\gamma,n+2h+\gamma],\omega).
\end{equation}
Notice that this is a double series, with $n$ dependence on the $\ \!_3F_2$. To understand the divergent behavior in $\omega$, as $\omega\rightarrow 1$, it needs to be resummed.

\subsubsection{Perturbative computation for the inhomogeneous (particular) solution}
The calculation is very similar to what we just did. We start with
 the  particular solution $\phi_{P}^{Tot}$ expanded as a perturbation series as follows
\begin{equation}
   \phi_{P}^{Tot}=\phi_{P}+\delta \phi_{P}
\end{equation}
where $\phi_{P}$ is the particular solution of the planar black hole given by \begin{equation*}
    \phi_{P}= \frac{B \omega^2}{(2-h) (h+1)}\sum^{\infty}_{n=0}j_{n}\omega^n, \quad j_{n}\equiv \frac{1}{n!}\frac{(1)_{n}(2)_{n}(2)_{n} }{(3-h)_{n}(h+2)_{n}}
\end{equation*}
The differential equation for the fir order  perturbation $\delta \phi_{p}$ is 
\begin{equation}\label{ODEPP}
 \omega^2(1-\omega)\delta \phi_{p}''-\omega^2\delta \phi_{p}'+h(1-h)\delta \phi_{p}=-s^2 \omega^{\gamma}(\gamma\omega \phi_{p}'+\omega^2 \phi_{p}'').
\end{equation}
Computing the source gives
\begin{equation*}
  \omega^2(1-\omega)\delta \phi_{P}''-\omega^2\delta \phi_{P}'+h(1-h)\delta \phi_{P}=-s^2 \frac{B }{(2-h) (h+1)} \sum^{\infty}_{n=0}\bar{J}_{n} \omega^{n+2+\gamma},
\end{equation*}
where we have defined the constant $\bar{J_{n}}$ as follows
\begin{equation}\label{Jn}
  \bar{J}_{n}\equiv\frac{1}{n!}\frac{(1)_{n}(2)_{n}(2)_{n} }{(3-h)_{n}(h+2)_{n}} (n+2)(n+1+\gamma).
\end{equation}
The solution we need is 
\begin{multline}
    \delta \phi_{P}=\frac{-B s^2   }{(2-h) (h+1)} \sum^{\infty}_{n=0}   \frac{\bar{J}_{n} \omega^{n+2+\gamma}}{(n+\gamma+2-h)(n+\gamma+h+1)}
    \\{}_{3}F_{2}([1,n+\gamma+2,n+\gamma+2],[n+\gamma+3-h,n+\gamma+2+h],\omega).
\end{multline}
Again, it should be noted that this is also a double series expansion, with $n$ dependence on the $\ \!_3F_2$, so one needs to do a resummation before one can understand the details of the divergent pieces. 

\subsubsection{Double limit series}

Now, instead of doing a resummation, we will consider the expression
\begin{equation}
  \delta \phi_{H}=-s^2 \sum^{\infty}_{n=0}\frac{\bar{A}_{n} \omega^{n+h+\gamma}}{(n+\gamma)(n+\gamma+2h-1)} {}_{3}F_{2}([1,n+h+\gamma,n+h+\gamma],[1+n+\gamma,n+2h+\gamma],\tilde\omega),\label{eq:newomega}
\end{equation}
where we have a new variable $\tilde \omega$. We want to take $\tilde\omega \to 1, \omega\to 1$ independently to understand the structure of the singularities. In the end we will make $\omega = \tilde \omega$ once we have cured the problems.
Because the double sum is not being done in standard form, this way of dividing the problem into two different variables might cause an order of limits problem. Essentially, if we take $\tilde \omega=\omega<1$, then the series is absolutely convergent and both ways of doing calculations give the same answer. Because there are  singularities in $\tilde \omega=1$, independent of $\omega$, when we take the double limit, the behavior can change. We will see that this happens in practice.

The idea is as follows. In \eqref{eq:newomega} we get a clear divergence in $\tilde \omega\to 1$ that we have already computed that is logarithmic in nature.
We find that
\begin{equation}
  \delta \phi_{H}=-s^2 \sum^{\infty}_{n=0}\bar{A}_{n} \omega^{n+h+\gamma} \frac{\Gamma(n+2h+\gamma-1)\Gamma(n+\gamma)}{\Gamma(n+h+\gamma)\Gamma(n+h+\gamma)} \log(1-\tilde \omega),\label{eq:newlimit}
\end{equation}
The sum over $n$ behaves as if the coefficients are approximately constant
\begin{equation}
   {\cal N} \sum_n \omega^n \sim \frac{{\cal N}}{(1-\omega)}
\end{equation}
times a normalization factor. This is because $a_n\sim 1/n$, the additional product of the Gamma functions also behaves exactly as $1/n$ when $n$ is large, and we're multiplying by a quadratic polynomial in $n$. The normalization factor was already found in \eqref{eq:logdiv1}
\begin{equation}
    {\cal N} = \frac{\Gamma(2h)}{\Gamma(h)^2}
\end{equation}

Doing a similar analysis for the other terms we find that 
\begin{eqnarray}
  \phi&=& \left[  c_1 \frac{ \Gamma (2 h) }{\Gamma (h)^2}\left(1 +s^2 \left(\mathcal{E}(d,h)+\frac{1}{1-\omega}\right) \right)+\right.
   \\ 
   &&\left.\frac{B_W (1-h)h \pi}{\sin(\pi h)}\left(1 +s^2\left(\mathcal{F}(d,h)+\frac{1}{1-\omega}\right)\right)\right]\log(1-\tilde \omega)
\end{eqnarray}
where 
\begin{eqnarray}
    \mathcal{E}(d,h)&\equiv& \lim_{{\omega}\to 1}\left[ h\Gamma(h)^2\Gamma\left(2h-1+\frac{2}{d}\right)\Gamma\left(\frac{2}{d}\right) \right.\nonumber
    \\ &&
   \left. {}_{4}\tilde{F}_{3}\left(\left[h,h+1,\frac{2}{d},2h+\frac{2}{d}-1\right],\left[2h,h+\frac{2}{d},h+\frac{2}{d}-1\right],\omega\right)-\frac{1}{1-\omega}\right]
    \label{E4f3},
\end{eqnarray}
and
\begin{eqnarray}
   \mathcal{F}(d,h)&\equiv&   \lim_{{\omega}\to 1}\left[\frac{-2\pi}{\sin{[\pi(h-2/d)]}} \frac{\Gamma\left(h+1+\frac{2}{d}\right)}{\Gamma \left(h-1-\frac{2}{d}\right)} \right.\nonumber
    \\ &
   {}_{5}\tilde{F}_{4}& \left.\left(\left[1,2,3,\frac{2}{d}+2-h,\frac{2}{d}+1+h\right],\left[3-h,h+2,\frac{2}{d}+2, \frac{2}{d}+1\right],\omega\right)
    -\frac{1}{1-\omega}\right]\label{F5f4}
\end{eqnarray}
are substracted versions of the functions that appear in our problem. The hypergeometric functions are regularized hypergeometric functions. It can be checked numerically that these regularized functions with the pole substracted actually have a limit. This indicates the absence of the double logarithmic divergence in our treatment of the sums.

It turns out that the single pole contribution that could in principle give rise to the double logarithmic behavior cancels between all the terms for the homogeneous problem by itself, which include the $h(1-h)$ term in the unperturbed differential operator, plus the expansion of the derivatives of the hypergeometric function on the source from the original problem. There are no such double logarithms generated in any intermediate step. This seems to be required so that the log divergences are unambiguous.

One can ask more generally, what is the source of the additional singular behavior? The idea is that the horizon is not longer at $\omega=1$, but that it has been shifted. The solution of the location of the horizon can be written as $\omega_h=1+\sum k_n s^{2n} $. The singularity is always logarithmic in terms of
\begin{equation}
    f(\omega) \log(\omega_h-\omega)
\end{equation}
as follows from the analysis of the singular points of the differential equation.
We can expand the logarithm as a power series in $s^2$, and we get
\begin{equation}
    \log(\omega_0-\omega)=\log(1+ \sum k_n s^{2n}-\omega)
    \sim \log(1-\omega)-\frac{\sum k_n s^{2n}}{1-w}+O(1/(1-\omega)^2)
\end{equation}
which gives an expansion in terms of a single logarithm and poles determined as a power series in $s^2$ by the shift of the horizon and the computation of $f$ to lower orders. There is no double logarithm ever.

Also, for non-extremal black holes is well-known that a field on a fixed background will be split into outgoing and infalling modes. These modes are mixed in the static limit $\nu\to 0$, and develop a logarithmic divergence 
\begin{equation*}
   C_{\pm}(\nu) (\omega_h-\omega)^{\pm i \nu}\approx \tilde{C}_{\pm}(\nu)( 1\pm i\log(\omega_h-\omega) )+O(\nu^2)
\end{equation*}
In this expression, $C_{\pm}(\nu)$ and $\tilde{C}_{\pm}(\nu)$ are constants with some dependence on the frequency $\nu$.  This logarithmic divergence basically comes from the coordinate singularity of the black hole. In fact, this expression is a remnant of Tortoise coordinate near the horizon $u^*=\frac{1}{2 \kappa} \log(u_h-u)$. From this point of view, the natural Boundary condition for the scalar field $\phi$ in any non-extremal black hole in static limit at the horizon is 
\begin{equation}\label{NaturalBoundaryCondition}
    \phi\approx A\log(\omega_h-\omega)
\end{equation}
where $A$ is a constant to determine from the exact solution of $\phi$ and where we have ignored some finite terms. In our perturbative treatment, we expand $A=A^{(0)}+s^2A^{(1)}+O(s^2)$ and $\omega_h=1+s^2+O(s^2)$, so the boundary condition becomes
\begin{equation}\label{ExpansionBD}
     \phi\approx A^{(0)}\log(1-\omega)+A^{(1)}s^2 \log(1-\omega)+\frac{ A^{(0)} s^2 }{1-\omega}+O(s^4)
\end{equation}
In this expansion in $s^2$, the boundary condition has picked an extra divergence coming from $\frac{ A^{(0)} s^2 }{1-\omega}$. This is an artifact of the perturbative treatment since the natural boundary condition is always (\ref{NaturalBoundaryCondition}). We can check explicitly that this expansion of the boundary condition in $s^2$ (\ref{ExpansionBD}) solves the differential equation (\ref{ODEspherical}) up to order $O(1-\omega)$ with the appropriate choice of $A^{(0)}$ given by $\frac{\Gamma(h)^2}{\Gamma(2h)}$ and $\frac{h(1-h)\pi}{\sin(\pi h)}$ for the homogeneous and particular solution respectively. We notice that the extra divergence is completely determined by the zeroth order solution which means that we can safely ignore it at order $s^2$.  Our task is reduced to find explicitly $A^{(1)}$ to find the leading correction to the thermal one-point function. This behavior is similar to analysis of the renormalization group, where higher order singularities are completely determined by the leading logarithmic singularity.

When we impose the first order result we already had, we see that the pole in $1-\omega$ cancels between the homogeneous solution and the inhomogeneous solution.
We are left with a strict logarithmic singularity at order $s^2$. A correction of order $s^2$ in the homogeneous solution can cancel this extra singularity.
The only possible source of this logarithmic singularity is the one that arises from the limits of the $\ {}_3F_2$ series that we computed. Hence, the calculation in this order is actually the correct calculation.

Notice also that the $\tilde F$ regularized hypergeometric functions are always well 
defined as functions of their parameters. There are additional poles that can appear in $\tilde F(d,h)$ arising from the $\sin(\pi(h-2/d))$ in the denominator. The poles occur when $h-2/d$ is an integer (except zero and one, where the $\Gamma(h-1-2/d)$  cancels them). These additional poles should also be interpreted as mixing, with operators that include the curvature of the background metric (which always has dimension $2$ and at least two powers of $:T:^2$).

The final result is 
\begin{equation}\label{eq:oneptsch}
  \left<O\right>_{J=0}=-N z_0^{-\Delta}B_W\frac{\Gamma (h)^2}{\Gamma (2 h)}\frac{\pi h(1-h)}{\sin(\pi h)}\left(1-s^2[\mathcal{F}(d,h)-\mathcal{E}(d,h)]\right)
\end{equation}
Notice that this is written in terms of $z_0$, which denotes the energy density. 

If one wants to use the temperature instead, one needs to notice that
that the temperature is given by a computation on the horizon
\begin{equation}
    T_{sbh}=\frac{d}{4\pi}\left(\frac{r_{h}}{L^2}\right)+\frac{d-2}{8\pi r_{h}}
\end{equation}
where $r_h$ is the radius of the horizon. In the large mass limit, 
the horizon is close to $r_h \sim r_0= L^2 z_0^{-1} $. The correction is small and the temperature of the black hole can be written as 
\begin{equation}\label{Leadingterm Tsb}
    T_{sbh}\approx T_{bb}\left(1+\frac{(d-3)}{d}\frac{L^2}{r_0^2}\right), \quad T_{bb}=\frac{d r_0}{4\pi L^2}
\end{equation}
where we compare to the black brane temperature. This introduces an additional correction of order $s^2$ in \eqref{eq:oneptsch}. 

\begin{equation}\label{eq:oneptschT}
  \left<O\right>_{J=0}=-N\left(\frac{4\pi T_{sbh}}{d}\right)^{\Delta}B_W\frac{\Gamma (h)^2}{\Gamma (2 h)}\frac{\pi h(1-h)}{\sin(\pi h)}\left(1-s^2[\mathcal{F}(d,h)-\mathcal{E}(d,h)]-s^2(d-3)h\right)
\end{equation}

\section{Conclusion}

The problem of linear response in AdS black holes and it's relation to conformal field theory correlators is an important problem on which a lot of recent progress has been made, especially for $AdS_5$ black holes. 

Our results, by contrast are valid for arbitrary dimension, but we are only able to make a perturbative analysis order by order. The zero-th order problem of q flat brane black holes at zero momentum is relatively simple.
Already the first non-trivial order is rather complicated. Our answers are given in terms of infinite sums of $\Gamma$ functions, which we showed could be matched to the $AdS_3$ setup, giving us confidence in our methods. 

Solutions in terms of other methods for $AdS_5$ give very complicated answers as well, in terms of solutions to the Heun differential equations. It would be interesting to understand the relation between these methods more carefully. 
Perhaps there is a way to simplify the final computations, which in our case get harder as we increase the order of the perturbation. 

More generally, we plan to study the problem of  retarded Green's function at arbitrary dimension, perturbing around the zero frequency, zero momentum
result.

In regards to the one point function in $AdS$ spherical black holes, we also found a complicated answer. It would be interesting if there is a better solution for the case of $AdS_5$ related to the other recent developments in the computation of retarded Green's functions, perhaps by understanding 
how solutions of Heun's equation with sources generate generalized solutions (similar to generalized hypergeometric functions). This is something we're currently looking into as well.  

Our method, which is perturbative, is probably also useful to study non-linear equations in the bulk, even at zero momentum. 
It would be interesting to also generalize our results beyond scalar fields.

On some other issues, like computing the time to the singularity by looking at the one point functions, as suggested in \cite{Grinberg:2020fdj}, we have nothing useful to say. We believe understanding this issue will require solving the exact equation in detail. The presence of additional singularities that can appear in the differential makes the analysis of such a problem a lot harder to undertake than in the flat brane case.

\acknowledgements
D.B. would like to thank G. Horowitz, J. Maldacena and J. Simon for various discussions. 
D.B.  research supported in part by the Department of Energy under Award No DE-SC 0011702.
R.M. research supported in part by the Department of Energy under Award No. DE-SC0019139


\begin{thebibliography}{99}





\bibitem{Maldacena:1997re}
J.~M.~Maldacena,
``The Large N limit of superconformal field theories and supergravity,''
Adv. Theor. Math. Phys. \textbf{2} (1998), 231-252
doi:10.1023/A:1026654312961
[arXiv:hep-th/9711200 [hep-th]].


\bibitem{Gubser:1998bc}
S.~S.~Gubser, I.~R.~Klebanov and A.~M.~Polyakov,
``Gauge theory correlators from noncritical string theory,''
Phys. Lett. B \textbf{428} (1998), 105-114
doi:10.1016/S0370-2693(98)00377-3
[arXiv:hep-th/9802109 [hep-th]].

\bibitem{Witten:1998qj}
E.~Witten,
``Anti-de Sitter space and holography,''
Adv. Theor. Math. Phys. \textbf{2} (1998), 253-291
doi:10.4310/ATMP.1998.v2.n2.a2
[arXiv:hep-th/9802150 [hep-th]].



\bibitem{Deutsch}
J.~M.~Deutsch, ``Quantum statistical mechanics in a closed system'',
Phys. Rev A, Vol 43, 4 (1991),  2046.
  doi: 10.1103/PhysRevA.43.2046

\bibitem{Srednicki:1994mfb}
M.~Srednicki,
``Chaos and Quantum Thermalization,''
doi:10.1103/PhysRevE.50.888
[arXiv:cond-mat/9403051 [cond-mat]].

\bibitem{Myers:2016wsu}
R.~C.~Myers, T.~Sierens and W.~Witczak-Krempa,
``A Holographic Model for Quantum Critical Responses,''
JHEP \textbf{05} (2016), 073
doi:10.1007/JHEP09(2016)066
[arXiv:1602.05599 [hep-th]].

\bibitem{Grinberg:2020fdj}
M.~Grinberg and J.~Maldacena,
``Proper time to the black hole singularity from thermal one-point functions,''
JHEP \textbf{03} (2021), 131
doi:10.1007/JHEP03(2021)131
[arXiv:2011.01004 [hep-th]].

\bibitem{Berenstein:2014cia}
D.~Berenstein and A.~Miller,
``Conformal perturbation theory, dimensional regularization, and AdS/CFT correspondence,''
Phys. Rev. D \textbf{90} (2014) no.8, 086011
doi:10.1103/PhysRevD.90.086011
[arXiv:1406.4142 [hep-th]].

\bibitem{Berenstein:2016avf}
D.~Berenstein and A.~Miller,
``Logarithmic enhancements in conformal perturbation theory and their real time interpretation,''
Int. J. Mod. Phys. A \textbf{35} (2020) no.29, 2050184
doi:10.1142/S0217751X20501845
[arXiv:1607.01922 [hep-th]].

\bibitem{Horowitz:2012ky}
G.~T.~Horowitz, J.~E.~Santos and D.~Tong,
``Optical Conductivity with Holographic Lattices,''
JHEP \textbf{07} (2012), 168
doi:10.1007/JHEP07(2012)168
[arXiv:1204.0519 [hep-th]].

\bibitem{Horowitz:2012gs}
G.~T.~Horowitz, J.~E.~Santos and D.~Tong,
``Further Evidence for Lattice-Induced Scaling,''
JHEP \textbf{11} (2012), 102
doi:10.1007/JHEP11(2012)102
[arXiv:1209.1098 [hep-th]].

\bibitem{Donos:2013eha}
A.~Donos and J.~P.~Gauntlett,
``Holographic Q-lattices,''
JHEP \textbf{04} (2014), 040
doi:10.1007/JHEP04(2014)040
[arXiv:1311.3292 [hep-th]].

\bibitem{Hartnoll:2014cua}
S.~A.~Hartnoll and J.~E.~Santos,
``Disordered horizons: Holography of randomly disordered fixed points,''
Phys. Rev. Lett. \textbf{112} (2014), 231601
doi:10.1103/PhysRevLett.112.231601
[arXiv:1402.0872 [hep-th]].

\bibitem{Son:2002sd}
D.~T.~Son and A.~O.~Starinets,
``Minkowski space correlators in AdS / CFT correspondence: Recipe and applications,''
JHEP \textbf{09} (2002), 042
doi:10.1088/1126-6708/2002/09/042
[arXiv:hep-th/0205051 [hep-th]].


\bibitem{Dodelson:2022yvn}
M.~Dodelson, A.~Grassi, C.~Iossa, D.~Panea Lichtig and A.~Zhiboedov,
``Holographic thermal correlators from supersymmetric instantons,''
[arXiv:2206.07720 [hep-th]].

\bibitem{Bhatta:2022wga}
A.~Bhatta and T.~Mandal,
``Exact thermal correlators of holographic $CFT$s,''
[arXiv:2211.02449 [hep-th]].


\bibitem{Bianchi:2021mft}
M.~Bianchi, D.~Consoli, A.~Grillo and J.~F.~Morales,
``More on the SW-QNM correspondence,''
JHEP \textbf{01} (2022), 024
doi:10.1007/JHEP01(2022)024
[arXiv:2109.09804 [hep-th]].




\bibitem{BarraganAmado:2018zpa}
J.~Barrag\'an Amado, B.~Carneiro Da Cunha and E.~Pallante,
``Scalar quasinormal modes of Kerr-AdS${_5}$,''
Phys. Rev. D \textbf{99} (2019) no.10, 105006
doi:10.1103/PhysRevD.99.105006
[arXiv:1812.08921 [hep-th]].


\bibitem{Aminov:2020yma}
G.~Aminov, A.~Grassi and Y.~Hatsuda,
``Black Hole Quasinormal Modes and Seiberg\textendash{}Witten Theory,''
Annales Henri Poincare \textbf{23} (2022) no.6, 1951-1977
doi:10.1007/s00023-021-01137-x
[arXiv:2006.06111 [hep-th]].


\bibitem{Bonelli:2021uvf}
G.~Bonelli, C.~Iossa, D.~P.~Lichtig and A.~Tanzini,
``Exact solution of Kerr black hole perturbations via CFT2 and instanton counting: Greybody factor, quasinormal modes, and Love numbers,''
Phys. Rev. D \textbf{105}, no.4, 044047 (2022)
doi:10.1103/PhysRevD.105.044047
[arXiv:2105.04483 [hep-th]].

\bibitem{Bonelli:2022ten}
G.~Bonelli, C.~Iossa, D.~Panea Lichtig and A.~Tanzini,
``Irregular Liouville Correlators and Connection Formulae for Heun Functions,''
Commun. Math. Phys. \textbf{397}, no.2, 635-727 (2023)
doi:10.1007/s00220-022-04497-5
[arXiv:2201.04491 [hep-th]].







\bibitem{Marolf:2004fy}
D.~Marolf,
``States and boundary terms: Subtleties of Lorentzian AdS / CFT,''
JHEP \textbf{05} (2005), 042
doi:10.1088/1126-6708/2005/05/042
[arXiv:hep-th/0412032 [hep-th]].

\bibitem{Berenstein:1998ij}
D.~E.~Berenstein, R.~Corrado, W.~Fischler and J.~M.~Maldacena,
``The Operator product expansion for Wilson loops and surfaces in the large N limit,''
Phys. Rev. D \textbf{59} (1999), 105023
doi:10.1103/PhysRevD.59.105023
[arXiv:hep-th/9809188 [hep-th]].



\bibitem{Blake:2019otz}
M.~Blake, R.~A.~Davison and D.~Vegh,
``Horizon constraints on holographic Green\textquoteright{}s functions,''
JHEP \textbf{01} (2020), 077
doi:10.1007/JHEP01(2020)077
[arXiv:1904.12883 [hep-th]].





\end{thebibliography}
\end{document}